\documentclass[aps,prl,twocolumn,showpacs,floatfix,amsmath,superscriptaddress]{revtex4}

\usepackage{bm}
\usepackage{graphicx}
\usepackage{latexsym}
\usepackage{amssymb}
\usepackage{color}

\begin{document}
\bibliographystyle{apsrev}

\title{Quantum magnetism and topological ordering via enhanced Rydberg-dressing near F\"orster-resonances}

\author{R.~M.~W.~van~Bijnen}
\affiliation{Max Planck Institute for the Physics of Complex Systems, N\"{o}thnitzer Strasse 38, 01187 Dresden, Germany}
\author{T.~Pohl}
\affiliation{Max Planck Institute for the Physics of Complex Systems, N\"{o}thnitzer Strasse 38, 01187 Dresden, Germany}

\date{\today}

\begin{abstract}
We devise a cold-atom approach to realizing a broad range of bi-linear quantum magnets. Our scheme is based on off-resonant single-photon excitation of Rydberg $P$-states (Rydberg-dressing), whose strong interactions are shown to yield controllable XYZ-interactions between effective spins, represented by different atomic ground states. The distinctive features of F\"orster-resonant Rydberg atom interactions are exploited to enhance the effectiveness of Rydberg-dressing and, thereby, yield large spin-interactions that greatly exceed corresponding decoherence rates. We illustrate the concept on a spin-1 chain implemented with cold Rubidium atoms, and demonstrate that this permits the dynamical preparation of topological magnetic phases. Generally, the described approach provides a viable route to exploring quantum magnetism with dynamically tuneable (an)isotropic interactions as well as variable space- and spin-dimensions in cold-atom experiments.
\end{abstract}


\maketitle
The exploration of quantum spin models is pivotal to the understanding of complex material properties \cite{auerbach} and has led to new concepts of phase transitions and discoveries of exotic quantum phases. A prominent example is the spin-1 Heisenberg chain, which forms the basis of Haldane's conjecture \cite{haldane83} about the emergence of a symmetry-protected topological phase \cite{pollmann12} that cannot be characterized by a local order parameter in terms of Landau's theory of phase transitions.

Cold atoms \cite{duan03,simon11,fukuhara13}, molecules \cite{gorshkov11,manmana13,yan13} and ions \cite{porras04,lanyon11,britton12,islam13,senko14} are emerging as promising platforms for a precise implementation of such spin models, leveraging on well controllable interactions and various lattice geometries with detailed experimental access to microscopic observables. This presents unique opportunities for studying exotic phases of quantum magnets and currently arising questions regarding their nonequilibrium physics. Yet, the realization of strong interactions and/or sufficiently low decoherence rates in order to observe long-time quantum dynamics and prepare low-energy, magnetic phases remains a major challenge. 

Here, we describe a cold-atom approach to a broad class of quantum spin models via optical dressing to high lying Rydberg states. By exploiting strong interactions and state mixing between highly excited states, we show how to realize various types of interactions between effective spins, represented by Rydberg-dressed ground states. We identify special configurations for which this Rydberg-dressing \cite{bouchoule02,henkel10,pupillo10, lauer12} yields enhanced interactions that greatly exceed associated decoherence rates.

\begin{figure}[t!]
\includegraphics[width = 0.99\columnwidth]{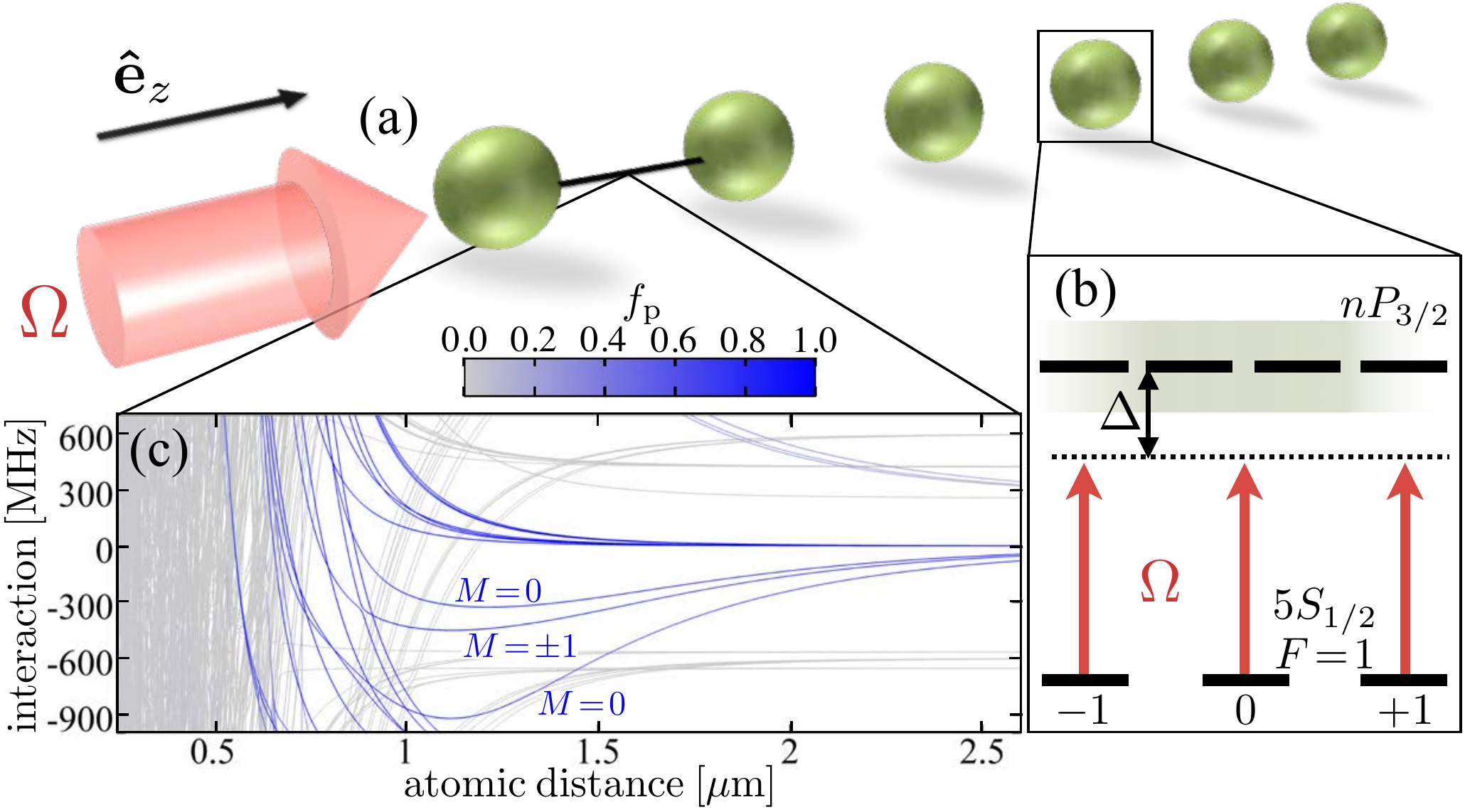}
\caption{\label{fig1} (color online) (a) A one-dimensional lattice of ultracold Rubidium atoms realizes an effective spin-1 chain, encoded in the $m_F=0,\pm1$ Zeeman states of the $F=1$ hyperfine manifold, as illustrated in panel (b). An elliptically polarized beam admixes excited states of a $nP_{3/2}$ Rydberg manifold with a detuning $\Delta$ and Rabi frequency $\Omega\ll |\Delta|$. The strong Rydberg interactions, shown in (c) for $n=43$, thereby induce effective interactions between the dressed ground states. Exciting close to minima of the interaction potential yields drastically enhanced interactions and permits to tune the relative strength of different interaction terms by selective coupling to molecular states with total angular momenta of $M=0$ or $M=\pm1$. The color coding indicates the fraction, $f_{\rm p}$, of optically coupled Rydberg-pair states for each molecular eigenstate \cite{suppl}.}
\end{figure}

\emph{Basic setup} -- As illustrated in Fig. \ref{fig1}a, we consider a linear chain of ultracold atoms confined within an optical lattice \cite{schauss12,schauss15} or an array of micro traps \cite{nogrette14,leung14}, with typical spacings of a few $\mu{\rm m}$ and temperatures of up to several  $10\mu$K \cite{nogrette14}. Focusing on the specific case of Rubidium atoms, the three $m_F=0,\pm1$ Zeeman states of the $|5S_{1/2},F=1\rangle$ ground state manifold provide a natural choice to encode an effective spin-1 degree of freedom (cf. Fig.  \ref{fig1}b). An elliptically polarised beam propagating along the chain axis couples these states to a Zeeman manifold of highly excited $nP_{3/2}$ Rydberg states with a large frequency detuning $\Delta$. The Rabi frequencies, $\Omega_{\pm}$, of its $\sigma_\pm$-polarization components are defined with respect to the $|m_F=\pm1\rangle\rightarrow|nP_{3/2},\mp1/2\rangle$ transition, with the total intensity scale $\Omega^2=\Omega_+^2+\epsilon^2 \Omega_-^2$ and ellipticity $\epsilon$. Due to the strong $\sim n^2$ scaling of their electronic dipole moments, the interaction between Rydberg states can exceed typical frequency detunings of the dressing laser. As a result, the strong Rydberg-Rydberg atom interaction dramatically alters the optical dressing of adjacent atom pairs, thereby leading to light-induced spin interactions. 

\emph{F\"orster resonant Rydberg interactions} -- At $\mu$m-scale distances, Rydberg-Rydberg atom interactions predominantly emerge from dipole-dipole coupling to other electronic pair states. Typically, the energy mismatch between two such pair states greatly exceeds their coupling strength, resulting in a van der Waals potential \cite{li05}. For certain Rydberg states, however, one finds accidental resonances of nearly degenerate pair states, at which interactions are greatly enhanced and acquire a longer ranged distance dependence \cite{li05}. Such so-called F\"orster resonances have been investigated in a number of Rydberg atom experiments \cite{vogt06,reinhard08,ditzhuijzen08,younge09,nipper12,ravets14} and were recently exploited \cite{tiarks14} for nonlinear optics applications \cite{gorshkov11b,gorniaczyk14}. 

\begin{figure}[t!]
\includegraphics[width = 0.99\columnwidth]{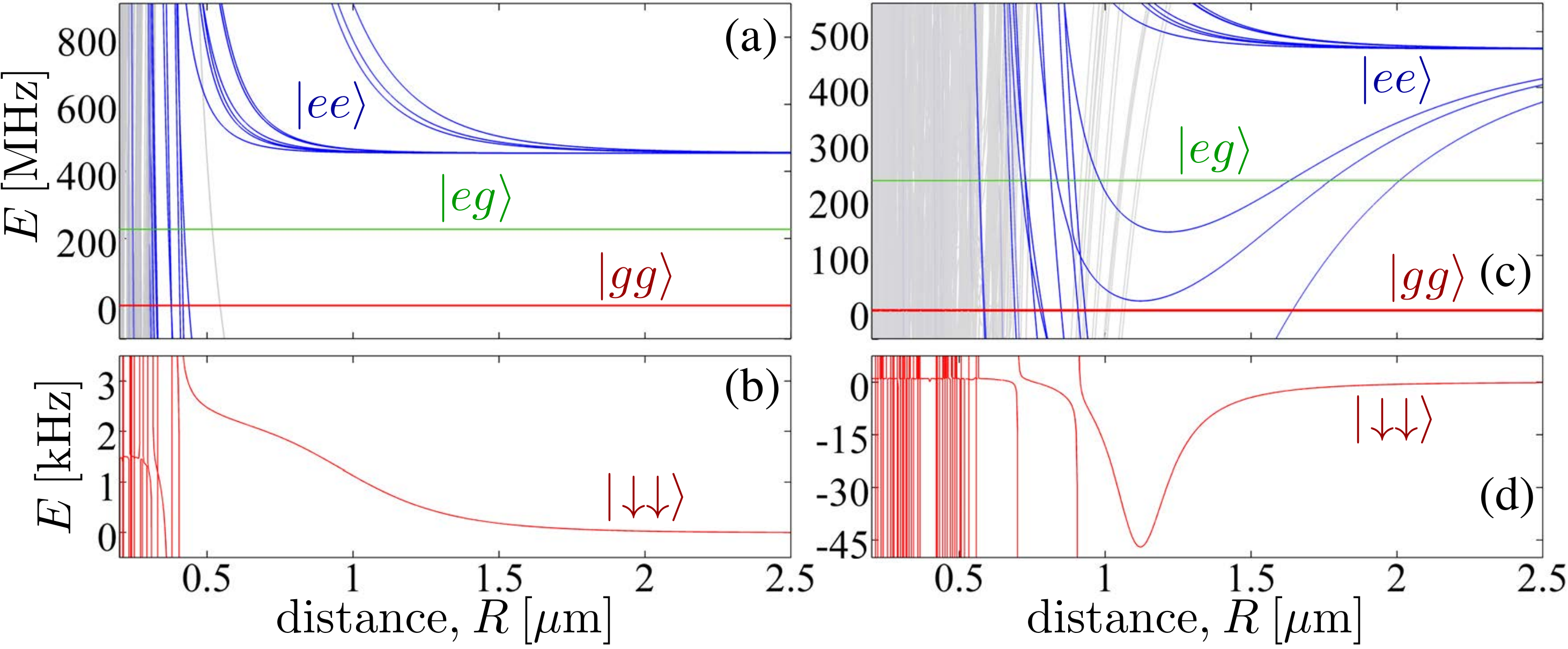}
\caption{\label{fig2} (color online) Pair potentials for optical excitation to (a,b) $32P_{3/2}$ and (c,d) $43P_{3/2}$ Rydberg states by $\sigma_+$-polarised light with $\Omega=10$MHz and $\Delta=-226$MHz. In both cases the maximum decay rate is $\bar{\gamma}/\gamma=5.5\times10^{-3}$, while the effective ground state interactions close to a F\"orster resonance (d) are strongly enhanced compared to purely repulsive interactions (b). The colors indicate the potentials of $2$-atom ground states ($|gg\rangle$) as well as singly ($|eg\rangle$) and doubly ($|ee\rangle$) excited states. Panels (b,d) show the detailed interaction energy for pairs of $|g\rangle=|\!\downarrow\rangle\equiv|5S_{1/2}m_J\!=\!-1/2\rangle$ fine structure ground states.}
\end{figure}

While recent experiments \cite{ravets14,beguin13,barredo14} with atoms at $R\gtrsim3\mu$m have confirmed this simplified description to a remarkable degree, it becomes increasingly inaccurate for smaller distances, $R\sim1\mu$m and $n\gtrsim 30$. We have hence performed numerical calculations of the molecular interaction potentials, using large basis sets of up to several $10^4$ states \cite{suppl}. Fig. \ref{fig1}c shows the resulting potential curves around an $(nP_{3/2}nP_{3/2})$-asymptote for $n=43$, slightly above a F\"orster resonance at $n=39$. This leads to asymptotically attractive interactions for $(nP_{3/2}m_J,nP_{3/2}m_J^\prime)$-pairs states with total angular momentum $M=m_J+m_J^\prime=0$ and $M=\pm1$. At smaller distances, these curves approach lower lying, repulsive $(nP_{1/2}nP_{3/2})$-pair potentials, and their strong dipole-dipole coupling leads to pronounced potential wells at $R_0\approx1\mu$m. Importantly, all intersecting molecular states are predominantly composed of $nP_J$ states with at least one $nP_{3/2}$-state atom, such that the formed wells still feature a large optical coupling strength for the ground$-$Rydberg state transition (cf. Fig.  \ref{fig1}c). Exploiting the vanishingly small hyperfine splitting of these Rydberg states, the effective spin interactions can be conveniently determined by first ignoring the hyperfine structure of the $5S_{1/2}$-states and subsequently projecting the resulting interactions between fine structure ground states, $|\!\!\uparrow\rangle=|5S_{1/2}m_J\!=\!1/2\rangle$ and $|\!\!\downarrow\rangle=|5S_{1/2}m_J\!=\!-1/2\rangle$, onto a given hyperfine $F$-manifold. 

\begin{figure}[b!]
\includegraphics[width = 0.99\columnwidth]{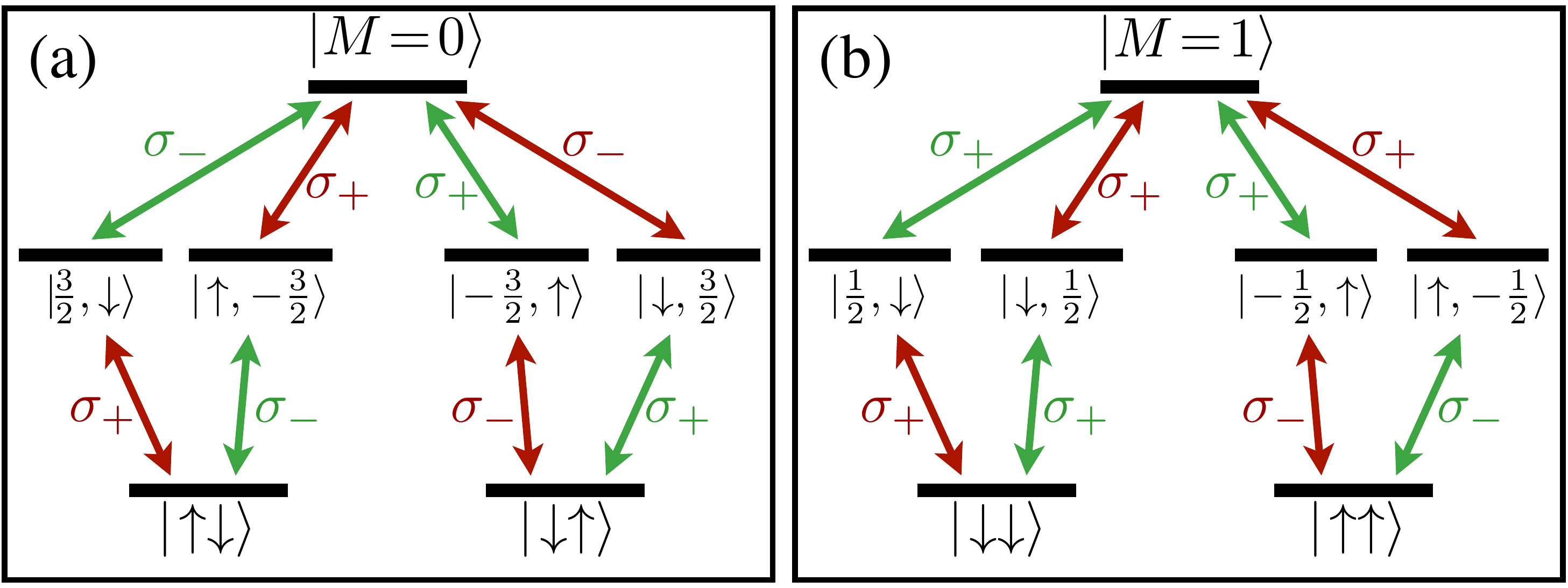}
\caption{\label{fig3} (color online) Examples of virtual four-photon processes leading to state-changing interactions between the fine structure ground states. The combination of $\sigma_+$ and $\sigma_-$polarized components opens up different interaction channels depending on the total angular momentum $M$ of the doubly excited Rydberg state.}
\end{figure}

\emph{Rydberg dressing} -- The major motivation behind the application of far-detuned light is to admix only a small Rydberg-state fraction, $P_{\rm ryd}\ll1$, to the dressed atomic ground state, such that it features a greatly suppressed decay rate $\bar{\gamma}=P_{\rm ryd}\gamma$ relative to that of bare Rydberg-state decay, $\gamma$ \cite{beterov09}. Existing approaches \cite{henkel10,pupillo10,maucher11,gil14,glaetzle14} are based on an excitation blockade \cite{lukin01}, which inhibits simultaneous dressing of two or more atoms within a critical distance $R_{\rm c}$ and gives rise to a soft-core potential, $\sim V_{\rm max}/(R^6+R_{\rm c}^6)$ \cite{henkel10}, arising from a distance-dependent two-body light shift $E(R)$. For a single ground and Rydberg state, the potential height, $V_{\rm max}$, corresponds to the differential light shift of an excitation-blocked ($R\rightarrow0$)and non-interacting atom pair ($R\rightarrow\infty$), $E(0)=-(\Delta-\sqrt{\Delta^2+2\Omega^2})/2$ and $E(\infty)=-(\Delta-\sqrt{\Delta^2+\Omega^2})$, respectively. Hence, $V_{\rm max}=E(0)-E(\infty)=-\Omega^4/(2\Delta)^3$ to leading order in the small parameter $\Omega/|\Delta|$. The small Rydberg state admixture $P_{\rm ryd}=\Omega^2/(2\Delta)^2$ yields a decoherence rate of $\bar{\gamma}=\gamma\Omega^2/(2\Delta)^2$. Thus the figure of merit, $\tfrac{|V_{\rm max}|}{\bar{\gamma}}=\tfrac{\Omega}{2|\Delta|}\tfrac{\Omega}{\gamma}$, for observing coherent quantum dynamics can be increased by maximizing the laser intensity but is always suppressed by $\Omega/|\Delta|\ll1$. Figs.~\ref{fig2}a and \ref{fig2}b show molecular potentials for dressing to $32 P_{3/2}$ states with $\sigma_+$ polarized light. While energy crossings with nearby pair states \cite{keating13} lead to increasingly dense resonances at smaller distances, the described plateau structure is still apparent for $R\gtrsim0.5\mu$m.

As shown in Figs.~\ref{fig2}c and \ref{fig2}d, the situation changes for F\"orster-resonant interactions where one can exploit the deep potential wells $U(R)$ to enhance the dressing-induced interactions. With the well depth $U_0=U(R_0)$ we define a total pair detuning $\delta=2\Delta-U_0$. Hence, $|\delta|$ can be made smaller than $|\Delta|$ to enhance the admixed fraction of Rydberg-pairs with respect to single excitations, thereby giving a more favourable ratio of interactions and decoherence rates. This can be seen from the two-body light shift at the potential well $E(R_0)=\tfrac{\Omega^2}{2\Delta}+\tfrac{\Omega^4}{4\Delta^2}(\delta^{-1}-\Delta^{-1})$, which gives $V_{\rm max}\approx\Omega^4/(U_0^2\delta)$ for $|\delta|\ll |\Delta|$. The excitation fraction $P_{\rm ryd}\approx\tfrac{\Omega^2}{U_0^2}(1+\Omega^2/\delta^2)$ is now determined by both single- and two-body excitations, such that $\tfrac{|V_{\rm max}|}{\bar{\gamma}}=\tfrac{\Omega/|\delta|}{1+(\Omega/\delta)^2}\tfrac{\Omega}{\gamma}$ assumes a maximum for $|\delta|=\Omega$. Hence, the corresponding ratio $\tfrac{|V_{\rm max}|}{\bar{\gamma}}=\tfrac{\Omega}{2\gamma}$ is strongly enhanced by a factor of $|\Delta|/ \Omega\gg1$, as shown in Fig. \ref{fig2}d.

\emph{Effective spin interactions} -- We now extend the above discussion to the simultaneous dressing of all four possible pair states ($|\!\!\uparrow\uparrow\rangle$, $|\!\!\downarrow\uparrow\rangle$, $|\!\!\uparrow\downarrow\rangle$, $|\!\!\downarrow\downarrow\rangle$) using van Vleck perturbation theory \cite{shavitt80} in the small parameter $\Omega / \Delta\ll1$ \cite{suppl}. Up to second order, the problem still factorizes and yields state-dependent light shifts $E_{\uparrow}$ and $E_{\downarrow}$, corresponding to an effective longitudinal magnetic field $E_{\uparrow}-E_{\downarrow}$. Interactions emerge within fourth order from distance-dependent energy shifts $E_{\uparrow\uparrow}(R)$, $E_{\downarrow\downarrow}(R)$ and $E_{\uparrow\downarrow}(R)=E_{\downarrow\uparrow}(R)$ of the four different pair states. Due to the different dipole matrix elements of the involved transitions between the two ground and the various Rydberg states these pair energies split. This gives rise to an effective Ising-type spin interaction $W_{zz}=E_{\uparrow\uparrow}-2E_{\uparrow\downarrow}+E_{\downarrow\downarrow}$ and an additional longitudinal magnetic field $(E_{\uparrow\uparrow}-E_{\downarrow\downarrow})/2$. 

\begin{figure}[t!]
\includegraphics[width = 0.99\columnwidth]{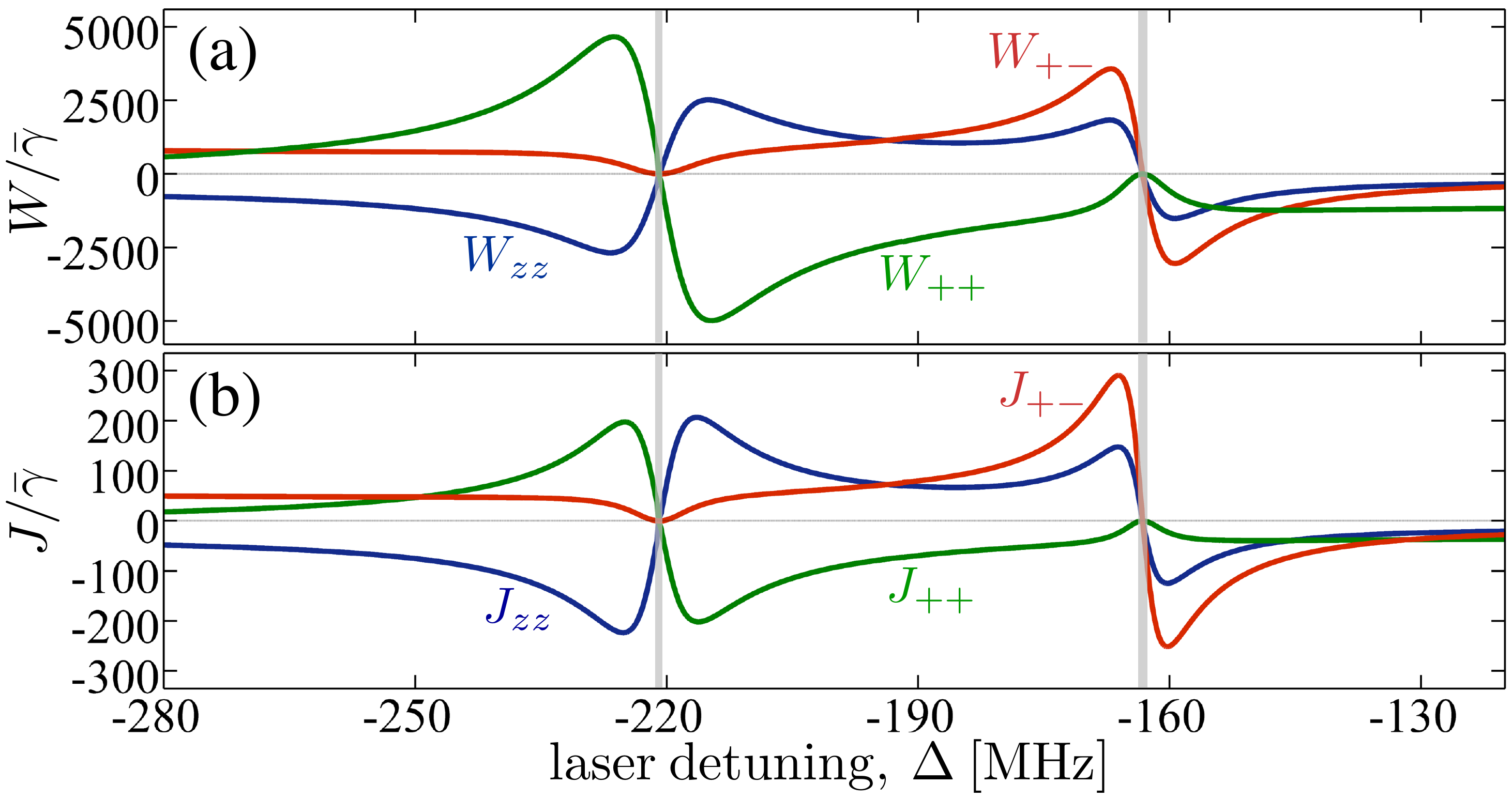}
\caption{\label{fig4} (color online) Strengths of dressing-induced interactions between (a) fine and (b) hyperfine ground states, scaled by the corresponding decoherence rate $\bar{\gamma}$. Rydberg $43P_{3/2}$ states are excited with $\Omega = 10 \mathrm{MHz}$ for an atomic distance $R = 1.2 \mu\mathrm{m}$. The interactions are obtained from 4th order perturbation theory which is accurate well outside of the grey-shaded regions corresponding to $|\delta|<\Omega^2/(2|\Delta|)$.}
\end{figure}

In addition to providing two-body level shifts, the Rydberg atom interactions also lead to strong Zeeman state mixing within a given Rydberg manifold \cite{walker08}. For the present geometry, each molecular Rydberg state is composed of all possible pair states $|nP_{3/2}m_J,nP_{3/2}m_J^\prime\rangle$ with a fixed total angular momentum $M=m_J+m_J^\prime$. For a finite ellipticity, $\epsilon>0$, the combination of $\sigma_+$- and $\sigma_-$-polarized light provides a two-photon coupling between the two-body ground and doubly excited Rydberg states with a total angular momentum change of $\Delta M=0,\pm2$. As illustrated in Fig. \ref{fig3}, this opens up two-body ground state transitions of the type $|\!\uparrow\downarrow\rangle\longleftrightarrow|\!\downarrow\uparrow\rangle$ ($\Delta M=0$) and $|\!\uparrow\uparrow\rangle\longleftrightarrow|\!\downarrow\downarrow\rangle$ ($\Delta M=\pm2$). The corresponding coupling strengths, $\sim\Omega^4/(U_0^2\delta)$, are comparable to the above Ising-type interactions, and their relative strength can be controlled by the laser detuning from the different potential wells. Exciting near the middle well (cf. Fig. \ref{fig1}c) with $M=\pm1$ predominantly dresses the $|\!\!\uparrow\uparrow\rangle$ and $|\!\!\downarrow\downarrow\rangle$ states and, therefore, yields a large coupling strength $W_{++}$ between these two pair states (cf. Fig. \ref{fig3}b). Analogously, tuning close to the upper well with $M=0$ provides enhanced flip-flop transitions between $|\!\uparrow\downarrow\rangle$ and $|\!\downarrow\uparrow\rangle$ (cf. Fig. \ref{fig3}a) with a frequency $W_{+-}$. 

\begin{figure}[b!]
\includegraphics[width = 1\columnwidth]{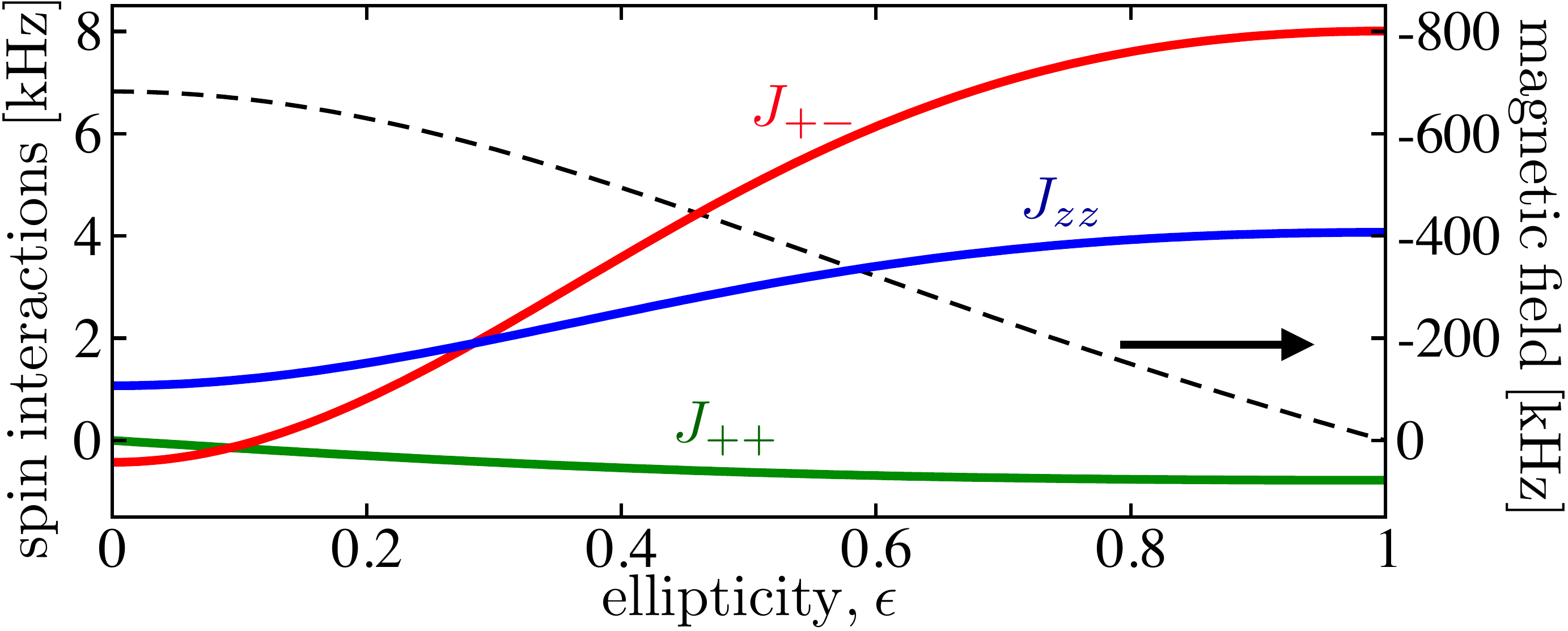}
\caption{\label{fig5} (color online) $\epsilon$-dependence of the interaction strengths and effective magnetic field for two atoms separated by $R=1.2\mu$m coupled to the $43P_{3/2}$ Rydberg manifold with  $\Omega = 10$ MHz and $\Delta = -163$ MHz.}
\end{figure}

Fig. \ref{fig4}a shows these interaction strengths as a function of the laser detuning covering the two uppermost potential wells in Fig. \ref{fig1}c. Indeed, the spin exchange interactions are widely tuneable, with dominant flip-flop interactions close to the $M=0$ resonance ($\Delta\approx-163$MHz) and vice versa at the $M=\pm1$ resonance ($\Delta\approx-221$MHz). Moreover, the laser detuning directly controls the sign of $W_{zz}$, giving antiferromagnetic interactions ($W_{zz}>0$) in between the two wells and ferromagnetic ones ($W_{zz}<0$) on the opposite side of the resonances.

In order to characterize the achievable degree of coherence we have scaled the interactions by the effective decoherence rate per atom, arising from spontaneous decay of the weakly admixed Rydberg states. As discussed above, the scaled interactions assume a maximum at $\delta\approx\Omega$ around each of the two resonances. Note that the depicted curves present an upper bound on decoherence effects, since we have used the maximum decoherence rate, $\bar{\gamma}_{\rm max}$, out of the four dressed pair states.  This implies that interactions can be made at least three orders of magnitude larger than the induced decoherence rates.

\begin{figure}[t!]
\includegraphics[width = 1\columnwidth]{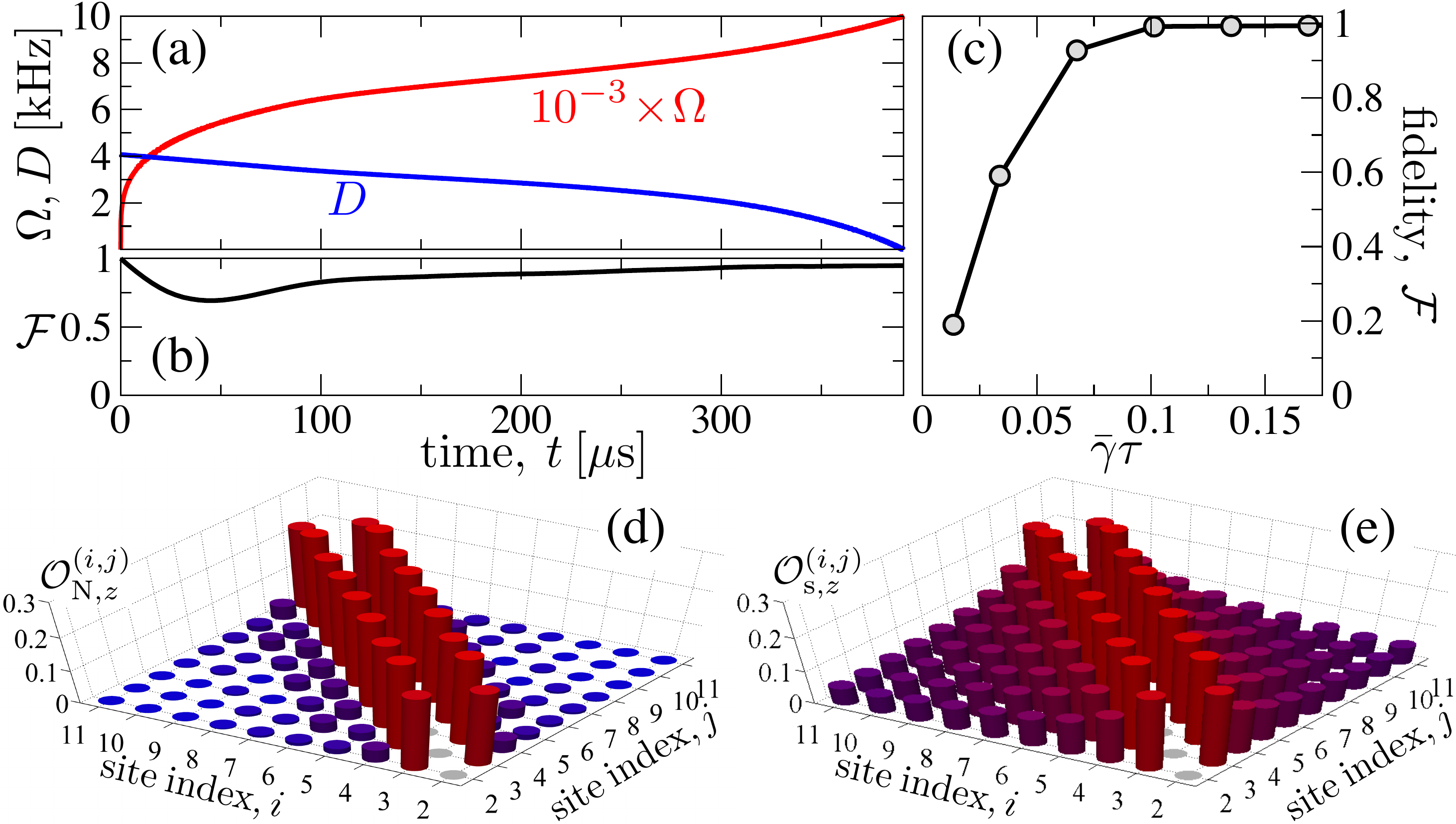}
\caption{\label{fig6} (color online) (a) Time evolution of $D$ and $\Omega$ used to steer the ground state of a 12-atom chain with a lattice constant of $1.2\mu$m, coupled to $43P_{3/2}$ Rydberg states with $\Delta=-163$MHz and $\epsilon=1$. (b) The large ground state fidelity indicates the high degree of adiabaticity for a short evolution time $\tau\approx0.4$ms. (c) $\tau$-dependence of the final ground state fidelity. (d,e) Short-range Neel-order and long-range string-order correlations, $\mathcal{O}_{{\rm N},z}^{(i,j)}$ (c) and $\mathcal{O}_{{\rm s},z}^{(i,j)}$ (d) of the final state, characteristic for the targeted Haldane phase. 
}
\end{figure}

\emph{Spin-1 Hamiltonian} -- Our perturbative treatment is accurate as long as $\Omega\!\ll\!|\Delta|$ and $|\delta|\gg\Omega^2/(2|\Delta|)$ (cf.\ Fig. \ref{fig4}). This limit permits a straightforward extension to $N$ dressed atoms, since the resulting many-body energy surfaces split into sums of $n$-particle interactions of order $\mathcal{O}[\varepsilon^{2(n-1)}]$ in the small parameter $\varepsilon=\Omega/\Delta$. Hence, it suffices to retain binary interactions, as derived above. For tight confinement, each singly occupied lattice site at position $X_i$ acts like a point-particle \cite{macri14} with the effective interaction strengths determined above. Finally, we incorporate the nuclear spin coupling by projecting the corresponding $N$-body Hamiltonian onto the $F=1$ ground state manifold, and obtain a general bilinear spin-1 XYZ-Hamiltonian of the form
\begin{eqnarray}\label{eq1}
\hat{H}&=&\sum_iB^{(i)}\hat{S}_z^{(i)}+D\left(\hat{S}_z^{(i)}\right)^2\nonumber\\
&&+\sum_{i<j}J_{zz}^{(i,j)}\hat{S}_z^{(i)}\hat{S}_z^{(j)}+J_{+-}^{(i,j)}\left(\hat{S}_x^{(i)}\hat{S}_x^{(j)}+\hat{S}_y^{(i)}\hat{S}_y^{(j)}\right)\nonumber\\
&&+\sum_{i<j}J_{++}^{(i,j)}\left(\hat{S}_+^{(i)}\hat{S}_+^{(j)}+\hat{S}_-^{(i)}\hat{S}_-^{(j)}\right)\;.
\end{eqnarray}
The spin-1 interactions $J_{zz}^{(i,j)}=W_{zz}(R_{ij})/16$, $J_{+-}^{(i,j)}=W_{+-}(R_{ij})/16$ and $J_{++}^{(i,j)}=W_{+-}(R_{ij})/8$ depend on the distances $R_{ij}=|X_i-X_j|$, but essentially provide nearest-neighbour interactions for lattice spacings $\sim R_0$. Despite their reduced strength, the interactions remain considerably larger than the corresponding decoherence rates (cf. Fig. \ref{fig4}b). The second term in Eq.~(\ref{eq1}) can be implemented via dressing to low-lying states with a linearly polarised field \cite{gerbier06}. The effective magnetic field $B^{(i)}=(E_{\downarrow}-E_{\uparrow})/4+\sum_{j\neq i}(E_{\downarrow\downarrow}(R_{ij})-2E_{\uparrow\downarrow}(R_{ij})+E_{\uparrow\uparrow}(R_{ij}))/8$ is dominated by second order light shifts and, hence, provides the largest energy scale. It can, however, be compensated via additional optical dressing to a low-lying state with circularly polarized light, or canceled entirely by choosing $\epsilon=1$ (cf. Fig. \ref{fig5}). 
In addition to tuning the magnetic field, the ellipticity, $\epsilon$, of the excitation laser also permits to tune the relative strength of the spin interactions while keeping $|\delta|\approx\Omega$. For $J_{++}\ll J_{+-},J_{zz}$ and $B^{(i)}=D=0$ the resulting XXZ-model undergoes a quantum phase transition from a Neel-ordered state to the Haldane phase at $J_{zz}\approx1.17 J_{+-}$ \cite{chen03}, which according to Fig. \ref{fig5} occurs at $\epsilon\approx0.3$.

The full dynamical control over the Hamiltonian (\ref{eq1}), e.g., enables the preparation of such topological phases in finite spin chains. Pumping all atoms into the $m_F=0$ state yields the many body ground state of Eq.~(\ref{eq1}) for vanishing interactions ($\Omega=0$) and $D>0$. Gradually reducing $D$ and simultaneously turning on the Rydberg coupling $\Omega$, thus, permits to transfer this initial state into the Haldane phase at $D=0$ and $0.85J_{+-}>J_{zz}>0$ (Fig. \ref{fig6}a). As shown in Fig. \ref{fig6}b for $N=12$ atoms, adiabaticity can be achieved for rather short evolution times during which just $\sim\!1$ of the atoms undergoes a single decay in average. The Haldane phase is characterized by nonlocal string order, $\mathcal{O}_{{\rm s},z}^{(i,j)}=\langle \hat{S}_z^{(i)}\prod_{k=i+1}^{j-1}e^{i\pi \hat{S}_z^{(k)}} \hat{S}_z^{(j)}\rangle$ \cite{nijs89}, in the absence of local order, $\mathcal{O}_{{\rm N},z}^{(i,j)}=\langle \hat{S}_z^{(i)}\hat{S}_z^{(j)}\rangle$. As shown in Figs.~\ref{fig6}c and d, the dynamically prepared many-body state indeed exhibits this behaviour to an extent that should be observable with current quantum gas microscope techniques, via state-selective imaging of the atoms \cite{endres11}.

\emph{Conclusion} -- We have described a cold-atom platform for a wide range of quantum magnets with strong spin interactions and long coherence times. While we have focussed here on naturally occurring F\"orster resonances, they can also be induced by external electric fields \cite{vogt06}, making our scheme applicable to other atomic species and higher principal quantum numbers, for which coherence can be increased even further. Dressing of either or both $F=1$ and $F=2$ hyperfine states permits to expand the underlying spin-dimension to $2$ or $3$ \cite{kjall13}. Alternatively, spin-$1/2$ models can be realized via selective dressing of two stable, energetically isolated states \cite{gil14,glaetzle14b}. In this case the present scheme yields larger interaction strengths (cf. Fig. \ref{fig4}a) that can exceed associated decoherence rates by more than 3 orders of magnitude. Moreover, the setting can be extended to higher dimensions and various lattice geometries, which opens the door to the rich physics of frustrated quantum magnets \cite{lhuillier01,lacroix11} including experimental searches for two-dimensional spin liquids \cite{balents10,yan11}. Here, the angular structure of the involved Rydberg states can also be used to engineer anisotropic interactions \cite{maucher11,glaetzle14,glaetzle14b}, where exciting below the demonstrated potential minima will prevent problems of resonant Rydberg excitation at otherwise occurring potential crossings. From a different perspective, our approach leads to a pronounced well structure of the state-dependent interactions (cf. Fig. \ref{fig2}d), which are sufficiently deep ($\sim\!\mu$K) to promote the formation of self-assembled quantum aggregates that provide an experimental platform to study quantum transport under controllable phonon coupling. 
The preparation of low-energy states of large lattices will naturally require a compromise between adiabaticity and decoherence processes. Yet, the tuneable coherence times and full dynamical control of the underlying spin model opens up promising ways to explore complex quantum evolution, including relaxation of nonlocal order \cite{mazza14} or the nonequilibrium physics of driven open quantum magnets \cite{lee13}. Rydberg dressing of two atoms has recently been demonstrated \cite{Jau15} and with the recent experimental breakthroughs \cite{schauss12,schauss15,ravets14,barredo14} in manipulating Rydberg states of atomic lattices all these applications may become accessible in the near future.

\begin{acknowledgments}We thank J. Zeiher, P. Schau\ss, C. Gross and I. Bloch for valuable discussions. This work was supported by the EU through the Marie Curie ITN "COHERENCE" and the EU-FET grant HAIRS 612862.

\emph{Note added} -- While writing this article we were informed by A. Glaetzle and P. Zoller about related work \cite{glaetzle14b} on 2D spin lattices using two-color Rydberg-dressing.
\end{acknowledgments}

\end{document}